\documentclass[iop]{emulateapj}
\usepackage{ulem}
\usepackage{url}
\usepackage{color}

\shorttitle{Sh 2-207}
\shortauthors{Yasui et al.}

\begin{document}

\title{Low-metallicity Young Clusters in the Outer Galaxy I. Sh 2-207} 
%\title{Metallicity Dependency of Protoplanetary Disk Lifetime for
%Intermediate-Mass Stars}

%\altaffilmark{1}}
%
%\altaffiltext{1}{Based on data collected at Subaru Telescope, which
%is operated by the National Astronomical Observatory of Japan.}

%%% \author{Genevieve J. Graves\altaffilmark{1,2,3} \&
%%%   S. M. Faber\altaffilmark{1}}
%%% 
%%% \altaffiltext{1}{UCO/Lick Observatory, Department of Astronomy and
%%% Astrophysics, University of California, Santa Cruz, CA 95064, USA}
%%% \altaffiltext{2}{Department of Astronomy, University of California,
%%% Berkeley, CA 94720, USA; graves@astro.berkeley.edu}
%%% \altaffiltext{3}{Miller Fellow}

\author{Chikako Yasui\altaffilmark{1, 2}, Naoto
Kobayashi\altaffilmark{2, 3, 4}, Alan T. Tokunaga\altaffilmark{5}, Masao
Saito\altaffilmark{6, 7}, and Natsuko Izumi\altaffilmark{2, 3}}
%, and Chihiro Tokoku\altaffilmark{??}} 
% \footnote{Also at: Subaru
%Telescope, National Astronomical Observatory of Japan, 650 North A`ohoku
%Place, Hilo, HI 96720, USA.}}
%
%\altaffiltext{3}{Miller Fellow}

\altaffiltext{1}{Department of Astronomy, Graduate School of Science,
University of Tokyo, Bunkyo-ku, Tokyo 113-0033, Japan}
%\affil{Institute of Astronomy, School of Science, University of Tokyo,
%2-21-1 Osawa, Mitaka, Tokyo 181-0015, Japan}
%
\email{ck.yasui@astron.s.u-tokyo.ac.jp}
%\email{ck.yasui@nao.ac.jp}
%\email{ck$_-$yasui@ioa.s.u-tokyo.ac.jp}

\altaffiltext{2}{Laboratory of Infrared High-resolution spectroscopy
(LIH), Koyama Astronomical Observatory, Kyoto Sangyo University,
Motoyama, Kamigamo, Kita-ku, Kyoto 603-8555, Japan}

\altaffiltext{3}{Institute of Astronomy, School of Science, University
of Tokyo, 2-21-1 Osawa, Mitaka, Tokyo 181-0015, Japan}
%\affil{Institute of Astronomy, School of Science, University of Tokyo,
%2-21-1 Osawa, Mitaka, Tokyo 181-0015, Japan}
%
%\email{ck$_-$yasui@ioa.s.u-tokyo.ac.jp}

\altaffiltext{4}{Kiso Observatory, Institute of Astronomy, School of
Science, University of Tokyo, 10762-30 Mitake, Kiso-machi, Kiso-gun,
Nagano 397-0101, Japan}

%\author{Alan Tokunaga\altaffilmark{4,5}}
%\author{Alan T. Tokunaga}
\altaffiltext{5}{Institute for Astronomy, University of Hawaii, 2680
Woodlawn Drive, Honolulu, HI 96822, USA}
%\affil{Institute for Astronomy, University of Hawaii, 2680 Woodlawn
%Drive, Honolulu, HI 96822, USA}

%\altaffiltext{6}{National Astronomical Observatory of Japan 2-21-1
%Osawa, Mitaka, Tokyo, 181-8588, Japan}
\altaffiltext{6}{Nobeyama Radio Observatory, 462-2 Nobeyama,
Minamimaki-mura, Minamisaku-gun, Nagano 384-1305, Japan}

\altaffiltext{7}{The Graduate University of Advanced Studies,
(SOKENDAI), 2-21-1 Osawa, Mitaka, Tokyo 181-8588, Japan}

%
%ALMA Project, National Astronomical Observatory of
%Japan, 2-21-1 Osawa, Mitaka, Tokyo 181-8588, Japan}

%\and

%\altaffiltext{4}{???}
%\author{Chihiro Tokoku} %\altaffilmark{4,5}}
%\altaffiltext{5}{Astronomical Institute, Tohoku University, Aramaki,
%Aoba, Sendai 980-8578, Japan}

%\affil{Astronomical Institute, Tohoku University,
%Aramaki, Aoba, Sendai 980-8578, Japan}

%%%%%%%%%%%%%%%%%%%%%%%%%%%%%%%%%%%%%%%%%%%%%%%%%%%%%%%%%%%%%%%%%%%%%%%%%%%%%%%

\begin{abstract}

To study star formation in low metallicity environments (${\rm [M/H]}
\sim -1$\,dex), we obtained deep near-infrared (NIR) images of Sh 2-207
(S207), which is an \ion{H}{2} region in the outer Galaxy with
spectroscopically determined metallicity of ${\rm [O/H]} \simeq
-0.8$\,dex.
We identified a young cluster in the western region of S207
with a limiting magnitude of $K_S =19.0$\,mag (10 $\sigma$) that
 corresponds to a mass detection limit of $\lesssim$0.1\,$M_\odot$ and
 enables the comparison of star-forming properties under low metallicity
 with those of the solar neighborhood. 
From the fitting of the K-band luminosity function (KLF), the age and
distance of S207 cluster are estimated at 2--3\,Myr and $\sim$4\,kpc,
respectively.
The estimated age is consistent with the suggestion of small extinctions
of stars in the cluster ($A_V \sim 3$\,mag) and the non-detection of
molecular clouds.
The reasonably good fit between observed KLF and model KLF suggests that
the underlying initial mass function (IMF) of the cluster down to the
detection limit is not significantly different from the typical IMFs in
the solar metallicity.
From the fraction of stars with NIR excesses, a low disk fraction
($<$10\,\%) in the cluster with relatively young age is suggested, as we
had previously proposed.
%

%  \begin{verbatim}
%  Abstract text is limited to 300 words.
%  A sentence acknowledging your funding agency is encouraged.
%  \end{verbatim}
% 

\end{abstract}

% http://www.journals.uchicago.edu/page/apj/instruct.key.html


% (Galaxy:) open clusters and associations: individual (..., ...) 

%%%%%%%%%%%%%%%%%%%%%%%%%%%%%%%%%%%%%%%%%%%%%%%%%%%%%%%%%%%%%%%%%%%%%%%%%%%%%%%
%%%%%%%%%%%%%%%%%%%%%%%%%%%%%%%%%%%%%%%%%%%%%%%%%%%%%%%%%%%%%%%%%%%%%%%%%%%%%%%
%%%%%%%%%%%%%%%%%%%%%%%%%%%%%%%%%%%%%%%%%%%%%%%%%%%%%%%%%%%%%%%%%%%%%%%%%%%%%%%

\section{INTRODUCTION} 
%%%%%%%%%%%%%%%%%%%%%%%%%%%%%%%%%%%%%%%%%%%%%%%%%%%%%%%%%%%%%%%%%%%%%%%%%%%%%%%
%%%%%%%%%%%%%%%%%%%%%%%%%%%%%%%%%%%%%%%%%%%%%%%%%%%%%%%%%%%%%%%%%%%%%%%%%%%%%%%
%%%%%%%%%%%%%%%%%%%%%%%%%%%%%%%%%%%%%%%%%%%%%%%%%%%%%%%%%%%%%%%%%%%%%%%%%%%%%%%
\label{sec:intro}

In a series of papers, we are going to present observations regarding
the properties of young low-metallicity clusters in order to
characterize the metallicity dependence of star-formation processes,
such as the initial mass function (IMF) and star formation efficiency.
The study of disk fraction for young clusters can also be used to 
characterize the metallicity dependence of planet formation processes
\citep{Yasui2009}.
However, the study of star formation under low metallicity is attracting
attention because it is important to characterize the star formation in
the early universe \citep{{Omukai2001},{Omukai2003}}. 
Moreover, it is critical for understanding the star formation process
itself.

Because of the Galactic metallicity gradient, our Galaxy contains
clusters with various metallicity ranging from ${\rm [M/H]} \sim -1.5$
in the outermost Galaxy to ${\rm [M/H]} \sim +0.5$ in the innermost
Galaxy. Our first targets are low-metallicity clusters because 1) the
large dynamic range of metallicity can be explored on the
low-metallicity side better than the high-metallicity side (up to ${\rm
[M/H]} \sim +0.5$);
2) low-metallicity are good alternative targets for studying star
formation in nearby dwarf galaxies that have similar low-metallicity
environments and much higher spatial resolution and sensitivity than the
substellar regime, and 3) they can serve as laboratories for studying
star formation in the epoch of Galaxy formation. 
Because the average metallicity in the universe is just {below} the
solar metallicity \citep{Matteucci2012}, clusters with subsolar
metallicity represent the most ubiquitous past star-formation.

For studying low-metallicity clusters, we focused on the outer Galaxy 
(Galactocentric radius: $R_{G} \gtrsim 15$\,kpc). 
Clusters in the outer Galaxy are much closer (heliocentric distance: 
$D \sim 10$\,kpc) than LMC/SMC ($D \sim 50$\,kpc), which is often
observed;
thus, they are well suited for precisely studying star formation in a
low-metallicity environment.
However, because the outer Galaxy is beyond the solar neighborhood
($D>2$\,kpc), we used near-infrared (NIR) deep $JHK_S$ images from the
Subaru 8.2 m telescope as primary sensitivity data down to the
substellar regime ($\lesssim$0.1\,$M_\odot$), which is similar to the
mass detection limit of extensive observations of embedded clusters in
the solar neighborhood with smaller (2--4\,m class) telescopes.
Large ground-based telescopes also offer high spatial resolution that
allows resolving stars in distant clusters.
Therefore, we can compare the properties of star formation in
low-metallicity environments with those in the solar neighborhood on the
same basis.

We selected sample clusters from the Sharpless catalog
\citep{Sharpless1959}, which lists photographically H$\alpha$-selected
bright \ion{H}{2} regions mostly in the 1st and 2nd quadrants of our
Galaxy. Although a number of lower mass clusters are known to exist in
the far outer Galaxy ($15\,{\rm kpc} < R_G < 18$\,kpc; e.g.,
\citealt{Snell2002}; \citealt{{Yun2007},{Yun2009}}) or the extreme outer
Galaxy ($R_{G} > 18$\,kpc; e.g., Cloud 2: \citealt{Kobayashi2000};
\citealt{Yasui2008}; WB89-789: \citealt{Brand2007}; Cloud 1:
\citealt{Izumi2014}), we focus on the relatively prominent young
clusters associated with Sharpless \ion{H}{2} regions to maximize the
astronomical properties of clusters with a significant number of cluster
members.

% Yun+2007:  IRAS 06361-0142 @Rg=15kpc, D=7kpc
% Yun+2009:  IRAS 07383-3325 @Rg=11.4kpc, D=5kpc
%x Palmeirim+2010: IRAS 07527-3446 @Rg=15.4kpc, D=10.3kpc

{As the first measurement of the metallicity and distance,} we refer to
\citet{Caplan2000} and \citet{Deharveng2000}. 
\citet{Deharveng2000} carefully assessed the abundance gradient of our
Galaxy with \ion{H}{2} regions using new observations \citep{Caplan2000}
%and Deharveng et all (2000), 
%who carefully assessed the abundance gradient of our Galaxy with
%\ion{H}{2} regions with their own new observations 
along with the data in the literature. 
A number of clusters were selected with known oxygen metallicity ${\rm
[O/H]} \leq -0.5$\,dex, assuming solar abundance of $12+\log {\rm (O/H)}
=8.73$ \citep{Asplund2009}.
We selected four of eight candidate clusters: Sh 2-207 (S207), Sh 2-208, 
Sh 2-209, as primary targets.

In this study, we present the first results for S207, which is one of
the lowest metallicity star-forming regions, $12 + \log ({\rm O/H})
\lesssim 8$ \citep{{Deharveng2000}, {Rudolph2006}}, in the Galaxy
located in the 2nd quadrant.
This paper is structured as follows: \S~2 discusses previous studies
about S207 and star-forming activities in S207 using WISE (Wide-field
Infrared Survey Explorer) and 2MASS (Two Micron All Sky Survey) data;
\S~3 describes Subaru MOIRCS (Multi-Object InfraRed Camera and
Spectrograph) deep $JHK_S$ images and data reduction; \S~4 describes the
results for a star-forming cluster in S207; and in \S~5, we discuss the
implications for the basic cluster parameters, such as age, distance,
IMF, and disk fraction.

%%%%%%%%%%%%%%%%%%%%%%%%%%%%%%%%%%%%%%%%%%%%%%%%%%%%%%%%%%%%%%%%%%%%%%
%%%%%%%%%%%%%%%%%%%%%%%%%%%%%%%%%%%%%%%%%%%%%%%%%%%%%%%%%%%%%%%%%%%%%%
\section{Sh 2-207} \label{sec:S207}
%%%%%%%%%%%%%%%%%%%%%%%%%%%%%%%%%%%%%%%%%%%%%%%%%%%%%%%%%%%%%%%%%%%%%%
%%%%%%%%%%%%%%%%%%%%%%%%%%%%%%%%%%%%%%%%%%%%%%%%%%%%%%%%%%%%%%%%%%%%%%

%X-ray
%Opt Halpha: IPHAS
%NIR 
%MIR/FIR: MSX/IRAS ??
%mm: CO ???

% General outline 
In this section, we discuss the properties of the target star-forming
region S207. 
In Table~\ref{tab:targets}, we summarize the properties from previous
works, i.e., the coordinate, distance, oxygen abundance, and
metallicity.
We also show NIR and mid-infrared (MIR) pseudocolor images of S207 in
Fig.~\ref{fig:3col_S207} (top). 
%Fig.~\ref{fig:3col_2MASS_WISE}.
% (see Section~\ref{sec:SFinS207} for more detail of
% Fig.~\ref{fig:3col_2MASS_WISE}).

\subsection{Basic properties from the literature} \label{sec:properties} 

S207 is located at {$l = 151.1905^\circ$ and $b = +2.1256^\circ$} on the
Galactic plane with coordinates $(\alpha_{\rm 2000.0}, \delta_{\rm
2000.0}) = (04^{\rm h} 19^{\rm m} 49.6^{\rm s}, +53^\circ 09' 29'')$
{from SIMBAD\footnote{This research has made use of the SIMBAD database,
operated at Centre de Donn\'ees Astronomiques de Strasbourg, France.} 
\citep{Wenger2000}}.
% 
%\citep{Beichman1988}.
% 04 19 49.6 +53 09 29
%\delta_{\rm 2000.0}) = (04^{\rm h} 19^{\rm m} 45^{\rm s}, +53^\circ 09'
%29'')$.  
%
It has an extended \ion{H}{2} region traced by H$\alpha$ emission 
\citep{Sharpless1959} and radio continuum emission
\citep{Condon1999}.
%; see Fig.~1 in \citealt{Fich1993} for S208\sout{xxx, see Figure 2}).
% 
It also accompanies strong MIR emission that is detected with IRAS, IRAS
04159+5302 {in IRAS Point Source Catalog \citep{Beichman1988} and
X0415+530 in IRAS Small Scale Structure} \citep{Helou1988}.
Despite the extensive search by \citet{Blitz1982}, no CO was detected
%\footnote{CO was detected at any of the positions we
%observed to the limit of our sensitivity (typically 0.5 K).}
with the limit of the sensitivity of $<$0.5\,K.
Therefore, S207 has been mistaken as a planetary nebula  (e.g.,
\citealt{Condon1999}).

The photometric distance, which is determined from spectroscopic and
photometric observations, is estimated at $\simeq$8\,kpc
for a probable dominantly-exciting O9.5IV type star\footnote{In
\citet{Crampton1978}, the spectral type is estimated to be O9V.},
% the brightest star in the S207 in visible, 
% (U=13.64, B=14, %V=13.23; REED+2003 AJ, 125, 2531-253)}
GSC 03719-00546 (7.6\,kpc by \citealt{Moffat1979} and 8.6\,kpc by
 \citealt{Chini1984}; see the large blue plus in {the top figure of}
 Fig.~\ref{fig:3col_S207}). 
%
%(8.6\,kpc, \citep{Chini1984} and 7.6\,kpc \citep{Moffat1979}
%$D_{\rm phot}=8.6$\,kpc \citep{Chini1984},
%$D_{\rm phot}=7.6$\,kpc \citep{Moffat1979}
%$D=8.6$\,kpc (Chini \& Wink 1984), 
%$D=7.6$\,kpc (Moffat 1979)
%
Assuming the Galactocentric distance of the Sun is $R_\odot = 8.0$\,kpc,
the distance corresponds to $R_G \simeq 15.5$\,kpc. 
However, note that the photometric distance is estimated at $D=1.8$\,kpc
($R_g \simeq 9.5$\,kpc) using the second brightest star in optical 
bands,
2MASS 04194732+5309216 (A7 III-type star; \citealt{Moffat1979}) shown by 
the small blue plus in 
%Fig.~\ref{fig:3col_2MASS_WISE}.
the top figure of Fig.~\ref{fig:3col_S207}.
%
%
%
%\noindent
%{\bi kinematic distance}: 
On the other hand, from observations of the H$\alpha$ radial velocity
using a Fabry--Perot spectrometer, the kinematic distance of S207 is
estimated at about 4\,kpc
(4.3\,kpc from $V_{\rm LSR}=-31.3$\,km\,s$^{-1}$ by Fich, Dahl, \&
 Treffers\ 1990; 3.4\,kpc from $V_{\rm LSR}=-35.4$\,km\,s$^{-1}$ by
 \citealt{Pismis1991}).
%\citealt{Fich1990}).
%$D_{\rm kin} = 3.4$\,kpc \citet{Pismis1991}
%(Pismis+ (1991) ,V(LSR)=-35.4 km/s),
%
%4.29\,kpc \citet{Fich1990}
%$D_{\rm kin} = 4.29$\,kpc (Fich, Treffers, \& Dahl (1990), V(LSR)=-31.3
%km/s),
%\bibitem[Fich et al.(1990)]{1990AJ.....99..622F} Fich, M., Dahl, G.~P., 
%\& Treffers, R.~R.\ 1990, \aj, 99, 622 
%
%$D\simeq 4$\,kpc ($R_g \simeq 11.7$\,kpc)
In this case, the distance corresponds to $R_g \simeq 12$\,kpc.
Considering the structure of our Galaxy, S207 is located beyond the
outer arm in the case of photometric distance,
whereas it is located within the outer arm in the case of kinematic
distance (e.g., Fig~1 of \citealt{Reid2014}).
Later, we will propose that the distance to this cluster is more likely
4\,kpc based on KLF analysis, which is similar to the above kinematic 
distance.

% {\bf Metallicity}
Based on the Fabry--Perot observations, \citet{Caplan2000} measured
several optical emission line fluxes in 36 \ion{H}{2} regions:
%fluxes of optical emission lines: 
%Caplan+ 2000: log(O/H) = 7.96(+0.17-0.27)
 [\ion{O}{2}] $\lambda\lambda$3726 and 3729, H$\beta$, [\ion{O}{3}]
 $\lambda$5007, [\ion{He}{1}] $\lambda$5876, and H$\alpha$.
Subsequently, \citet{Deharveng2000} derived the oxygen abundance (O/H),
as well as the extinctions, electron densities and temperatures, and
ionic abundances (O$^+$/H$^+$, O$^{++}$/H$^+$, and He$^+$/H$^+$). 
The estimated oxygen abundance of S207 is $12 + \log ({\rm O/H}) =
 8.02$. 
\citet{Rudolph2006} reanalyzed the elemental abundances of 117
\ion{H}{2} regions with updated physical parameters.
Among them, the oxygen abundance of S207 is estimated using the data of
\citet{Caplan2000} to be $12 + \log ({\rm O/H}) = 7.96^{+0.17}_{-0.27}$.
This corresponds to the metallicity of ${\rm [O/H]} \simeq -0.8$\,dex
assuming solar abundance of $12 + \log ({\rm O/H}) = 8.73$
\citep{Asplund2009}.
The spatial distribution of the Galactic abundance using the
spectroscopy of Cepheids \citep{Luck2006} also suggests low 
metallicity ($\lesssim$$-$0.5\,dex) in the outer Galaxy with distance of
$\gtrsim$4\,kpc in the second quadrant, where S207 is located.

\vspace{1em}

\subsection{Star-forming activities} \label{sec:SFinS207} 

Before discussing the results of our deep NIR imaging with Subaru,
we discuss the star-forming activities in S207 with 2MASS
\citep{Skrutskie2006} NIR data and WISE \citep{Wright2010} MIR data. 
The top figure of Fig.~\ref{fig:3col_S207} shows a pseudocolor image of
S207 with a wide field of view ($\sim$10$'\times$10$'$) with the center
at 
%$(l, b) = (151.19^\circ, +2.13^\circ$) in the Galactic coordinates.
$(l, b) = (151.1905^\circ, +2.1256^\circ$) in Galactic coordinates and
$(\alpha_{\rm 2000.0}, \delta_{\rm 2000.0}) = ( 04^{\rm h} 19^{\rm m}
49.6^{\rm s}, +53^\circ 09' 29'')$ in Equatorial coordinates.
%
%top image of Fig.~\ref{fig:3col_2MASS_WISE} 
The figure is produced by combining the
% MOIRCS $K_S$ (2.15\,$\mu$m)-band (blue), 
2MASS Ks-band (2.16\,$\mu$m, blue), WISE band 1 (3.4\,$\mu$m; green),
and WISE band 3 (12\,$\mu$m; red) images.
We also show the 1.4\,GHz radio continuum by the NRAO VLA Sky Survey (NVSS;
\citealt{Condon1998}) with white contours, which are the same as the 
contours in Fig.~1 in \cite{Condon1999} for IRAS04159+5302. 
{The 12\,$\mu$m emission is mainly from PAH emission}, tracing
photodissociation regions around \ion{H}{2} regions, whereas the radio
continuum traces the {photoionized} \ion{H}{2} region. 
The distributions of the 12\,$\mu$m emission and radio continuum show
that the \ion{H}{2} region extends almost spherically with an
approximately 2.5\,arcmin radius.
From the nearly perfect spherical shape centered on GSC 03719-00546, the
O-type star should be the exciting source of the \ion{H}{2} region. 
Although there is a strong radio continuum feature located $\sim$2$'$
toward the south-west of the star, this is likely to be a background
object, such as a quasar, because no NIR counterpart is seen.

\color[named]{Gray}
\color{black}

\section{OBSERVATION AND DATA REDUCTION} \label{sec:obs}

\subsection{Subaru MOIRCS imaging} \label{sec:obs_MOIRCS}
Because our targets are distant clusters ($D>2$\,kpc) beyond the solar
neighborhood, we used NIR deep $JHK_S$ images taken with the Subaru
8.2\,m telescope as primary data with sensitivity down to the substellar
regime and high spatial resolution to resolve stars in those distant
clusters. NIR is currently best suited for stellar counting of distant
young clusters compared to MIR.

%\noindent
%{\bf Subaru MOIRCS Imaging} \\
% \label{subsec:MOIRCS}
Deep $JHK_S$-band images were obtained for each band with the 8.2\,m
Subaru telescope equipped with a wide-field NIR camera and spectrograph,
MOIRCS \citep{{Ichikawa2006},{Suzuki2008}}.
%\citep[Multi-Object InfraRed Camera and Spectrograph;
%][]{{Ichikawa2006},{Suzuki2008}}.
%
MOIRCS used two 2K ``HAWAII-2'' imaging arrays, which yield a $4' \times
7'$ field of view ($3.5' \times 4'$ for each chip) with a pixel scale of
$0''.117$ pixel$^{-1}$.
The instrument uses the Mauna Kea Observatory (MKO) NIR photometric
filters \citep{{Simons2002},{Tokunaga2002}}.

The observations were performed on 2006 November 8 UT, when it was
highly humid ($\sim$45--75\,\%). The seeing was $\sim$1$''$.
Because the detector output linearity is not guaranteed for counts over
$\sim$20,000\,ADU, we obtained short exposure images 
% for avoiding saturation 
% obtain additional short exposure J and H band images 
in addition to long-exposure images for more sensitive detection.
The exposure times for the long-exposure images are 120, 20, and 30\,s
for the J, H, and $K_S$ bands, respectively, whereas the exposure time 
for short-exposure images is 13\,s for all bands.
The total integration times for the long-exposure images are 360, 480,
and 960\,s for the J, H, and $K_S$ bands, respectively, whereas the
total integration time for the short-exposure images is 52\,s for all
bands.
The center of the images of S207 is set at $\alpha_{\rm 2000} = 04^{\rm
h} 19^{\rm m} 56^{\rm s}$, 
{$\delta_{\rm 2000} = +53^\circ 09' 33''$}, which covers the whole
\ion{H}{2} region described in Section~\ref{sec:SFinS207} (see white box
in {the top figure of} Fig.~\ref{fig:3col_S207} for the MOIRCS field).
For background subtraction, we also obtained images of the sky, which is
4\,arcmin south of S207, to avoid the nebulosity of S207.
We summarize the details of the observation in Table~\ref{tab:LOG}.

%%% 
%%% 
%%% \clearpage
%%% \begin{verbatim}
%%% Name	D (kpc)	Rg (kpc)	[O/H]	観測モード	short	long
%%% Digel Cloud 2	12	19	-0.5〜-0.8	self sky	IRCS	2006.09.02
%%% Sh 2-209	8.6	17	-0.6	obj+sky	2006.11.07 (2008.01.12/13)	2006.09.02
%%% Sh 2-207	D	16.8	-0.9	obj + sky	2006.11.07	2006.11.07
%%% Sh 2-208	D	16.8	-1.0	obj + sky	2006.11.07
%%% 2007.11 (サービス, HK, Jは途中まで) / 2008.01 (J & Ks, メインfi
%%% \end{verbatim}
%%% 

%\noindent
\subsection{Data Reduction} \label{sec:Data}
%{\bf Data Reduction} \\
All data in each band were reduced using IRAF\footnote{IRAF is
distributed by the National Optical Astronomy Observatories, which are
operated by the Association of Universities for Research in Astronomy,
Inc., under cooperative agreement with the National Science Foundation.}
with standard procedures, including flat fielding, bad-pixel correction,
median-sky subtraction, image shifts with dithering offsets, and image
combination.
%
% \textcolor[named]{Gray}{\sout{We found that images after flat fielding using
% dome flat have global gradient with $\lesssim$5\,\%, fainter around the
% edge of images.
%
% The plate for dome flat image may not be illuminated uniformly due to
% the wide FOV of MOIRCS.}}
%
%Because a photometric accuracy of $\sim$0.01\,mag is necessary for
%deriving disk fraction, 
We used sky flats, which are made from 
%data obtained during the close run 
%
%For the data in 2006, we used a sky flat kindly provided by the MOIRCS
%support astronomer, Dr. Ichi Tanaka, while for data in 2007/2008, 
%We made sky flats using 
the archived MOIRCS data in SMOKA\footnote{SMOKA is the
Subaru--Mitaka--Okayama--Kiso Archive System operated by the Astronomy Data
Center, National Astronomical Observatory of Japan)}. We selected the
data of the closest run. 
In addition to the above standard procedures, distortion correction was
applied before image combination using the 
``MCSRED''\footnote{\url{http://www.naoj.org/staff/ichi/MCSRED/mcsred\_e.html}} 
reduction package for the MOIRCS imaging data. 
We constructed a pseudocolor image of S207 by combining the
long-exposure images for $J$ (1.26\,$\mu$m, blue), $H$ (1.64\,$\mu$m,
green), and $K_S$ (2.14\,$\mu$m, red) bands (Fig.~\ref{fig:3col_S207}{,
bottom}).

%http://www.naoj.org/staff/ichi/MCSRED/mcsred_e.html

\subsection{Photometry}  \label{sec:phot}
%\noindent
%{\bf Photometry and Limiting Magnitudes}\\
%
$JHK$ photometry was performed using the IRAF apphot package for stars
in the western half-frame of the image for S207 (``S207 frame,''
hereafter), where a star-forming cluster was found
(Fig.~\ref{fig:3col_S207}{, bottom}; see
Section~\ref{sec:S207cluster}). 
For comparison, we also performed photometry of stars in the western
half{-}frame of the image of the sky (``sky frame,'' hereafter).
As photometric standards, {isolated} 2MASS stars in these fields were
used after converting the 2MASS magnitudes to the MKO magnitudes using
the color transformations in \citet{Leggett2006}.
Only 2MASS stars {with good 2MASS photometric quality and} with colors
of $J-H \le 0.5$, $H-K_S \le 0.5$, and $J-K_S \le 1.0$ {were used} to
avoid the color term effect.
 The zero point magnitudes both for the S207 frame and the sky frame
were decided with relatively small dispersions: $\le$0.08\,mag for J and
H bands, and $\le$0.05\,mag for Ks bands.
% 0.07 mag, and 0.05 mag, respectively, for the sky frame, while 0.07mag,
% 0.08 mag, and 0.04 mag, respectively, for the sky frame. 
The difference of zero point magnitudes between two frames are very
 small, 0.03\,mag, 0.005\,mag, and 0.02\,mag for J, H, and K bands,
 respectively, confirming that the photometry for both frames is
 consistent. 
Because the cluster is very crowded, we used aperture diameter of
$0''.7$ for cluster members to avoid the contamination of adjacent
stars.
The flux uncertainty in the $0''.7$ aperture was estimated for each
frame from the standard deviation of the flux in about 6000 independent
apertures in the blank area in the frame.
% Compared to the flux uncertainty estimated by APPHOT from
%pixel-to-pixel variation, $\sigma$ is larger by a factor of 1.73, 1.73,
%and 1.64 for J, H, and Ks bands, respectively.  

The limiting magnitudes (10$\sigma$/5$\sigma$) of long-exposure images
for the S207 frame are $J=19.6/20.4$\,mag, $H=19.1/20.0$\,mag, and
$K_S=19.0/20.0$\,mag.
For the sky frame, the limiting magnitude in H-band is slightly brighter
compared to the S207 frame, $H=19.0/19.8$\,mag, while those in the J and
$K_S$ bands are the same as those for the S207 frame
%$J=19.6/20.4$\,mag, $H=19.0/19.8$\,mag, and $K_S=19.0/20.0$\,mag} 
(see Table~\ref{tab:limit}).
For J and H bands for the S207 frame and the sky frame, the differences
of limiting magnitudes between two frames are consistent with the
differences of the exposure times.
The limiting magnitudes in $K_S$ band are the same for both frames
despite the different exposure times, 1.3 times longer for the S207
frame,
probably owing to the nebulosity in the S207 frame, which is most
prominent in $K_S$ band (see Fig. 1).
Because there are no significant differences in the limiting magnitudes
between the S207 frame and the sky frame, we can directly compare the
number of stars in the S207 frame and those in the sky frame.

\section{A newly identified young cluster in S207}
\label{sec:results} 

\subsection{Identification of a Young Cluster in S207}
\label{sec:S207cluster}

\begin{figure}[h]
\begin{center}
%\includegraphics[width=13cm]{/Users/chikako/WORK/MOIRCS2006/S207/fitsfile_WORK/ds9_2.ps}
%\includegraphics[width=12cm]{/Users/chikako/WORK/MOIRCS2006/S207/fitsfile_WORK/ds9_0917.ps}
%
% /Users/chikako/WORK/MOIRCS2006/S207/fitsfile_WORK\/ds9_621.ps
% \includegraphics[width=18cm]{fig2.eps}
%
%
%\includegraphics[width=13cm]{fig1.eps}
%\includegraphics[width=17cm]{ds9_S207_fig1ab.eps}
\includegraphics[width=16cm]{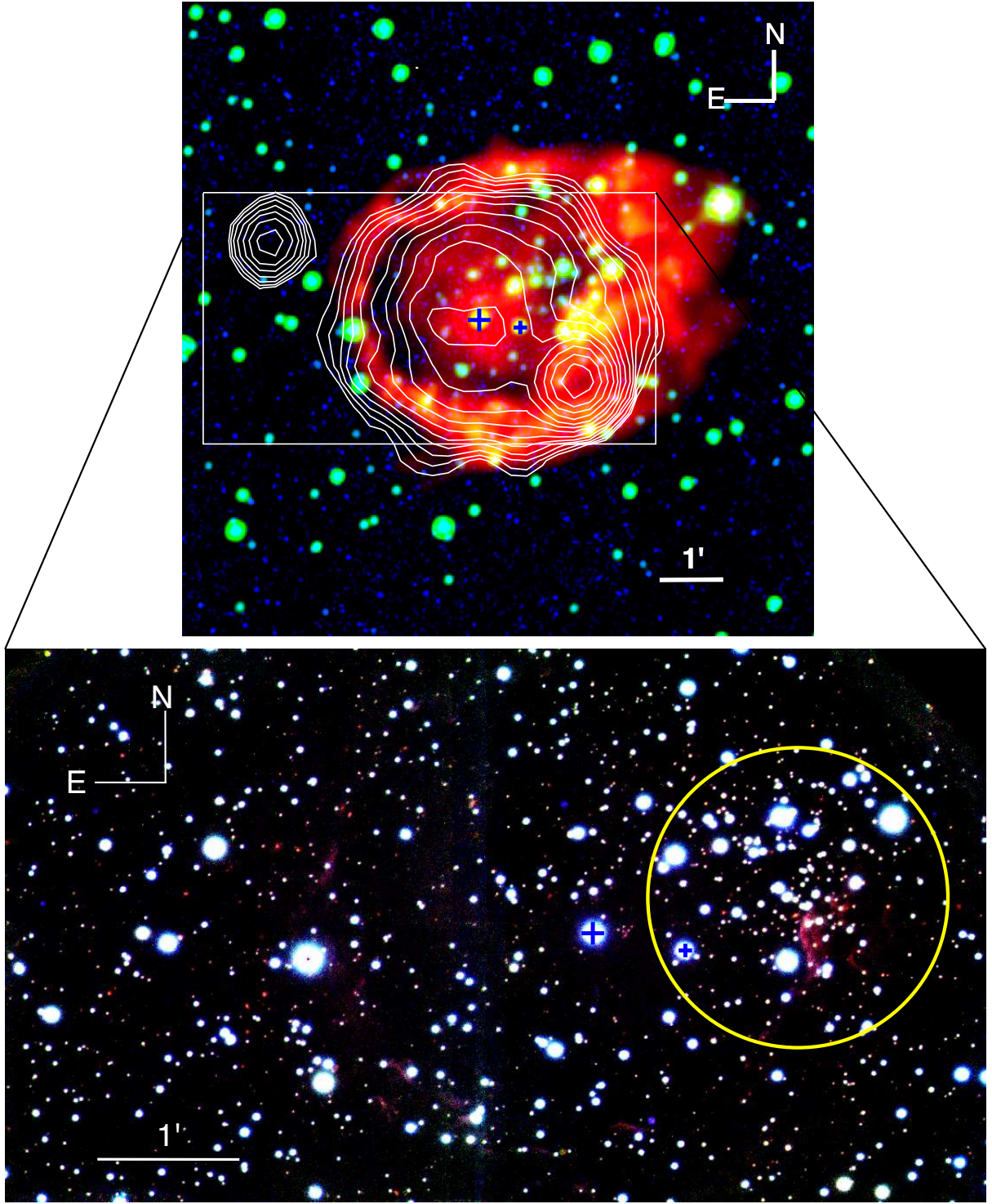}
\caption{Pseudocolor image of S207{. (Top) The image is produced by
 combining the 2MASS Ks-band (2.16\,$\mu$m; blue), WISE band 1
 (3.4\,$\mu$m; green), and WISE band 3 (12\,$\mu$m; red)}
%
%Pseudocolor image of S207
%. The field of view of the image is
% $\sim$10$'\times$10$'$ 
with a wide field of view ($\sim$10$'\times$10$'$) and the center of
{$(l, b) = (151.1905^\circ, +2.1256^\circ$)} in Galactic coordinates 
%04h19m50.70s +53d09m41.18s
and $(\alpha_{\rm 2000.0}, \delta_{\rm 2000.0}) = ( 04^{\rm h} 19^{\rm
 m} 49.6^{\rm s}, +53^\circ 09' 29'')$ in Equatorial coordinates.
%04^{\rm h} 19^{\rm m} 49.6^{\rm s}, +53^\circ 09' 29'')$
%
The 1\,arcmin corresponds to 2.4\,pc and 1.2\,pc for distances of S207
 of 8\,kpc and 4\,kpc, respectively.
The 1.4\,GHz radio continuum emission by NVSS is also shown using the
 white contours.  The contours are plotted at 1\,mJy\,${\rm beam}^{-1} 
 \times 2^0, 2^{-1/2}, 2^1$, ... . 
The blue plus symbols show the two brightest stars in the optical bands;
the larger symbol denotes the most brightest star (GSC 03719-00546), and
the smaller symbol denotes the second brightest star (2MASS
04194732+5309216; see details in the main text). 
The white box shows the location and size of the MOIRCS field of
 view.  
(Bottom) $JHK_S$ pseudocolor image of S207.  
The color image is produced by combining the $J$- (1.26\,$\mu$m), $H$-
(1.64\,$\mu$m), and $K_S$-band (2.14\,$\mu$m) images obtained with
MOIRCS at the Subaru telescope on November 2008 with the center of
$(\alpha_{\rm 2000.0}, \delta_{\rm 2000.0}) = ( 04^{\rm h} 19^{\rm m}
56^{\rm s}, +53^\circ 09' 33'')$ in Equatorial coordinates. 
The field of view of the  image is $\sim$7$'\times$4$'$.
The yellow circle ($r=64''$) shows the location of the cluster with
central coordinate of $(\alpha_{\rm 2000}, \delta_{\rm 2000}) = (04^{\rm
h} 19^{\rm m} 42.0^{\rm s}, +53^\circ 09' 43.9'')$. 
 The blue plus symbols show the same stars as in the top
 figure.} 
%\label{fig:3col_2MASS_WISE} 
\label{fig:3col_S207}
% \label{fig:3col_WISE}   
%
\end{center}
\end{figure}

\begin{figure}[!h]
\begin{center}
\includegraphics[width=8.5cm]{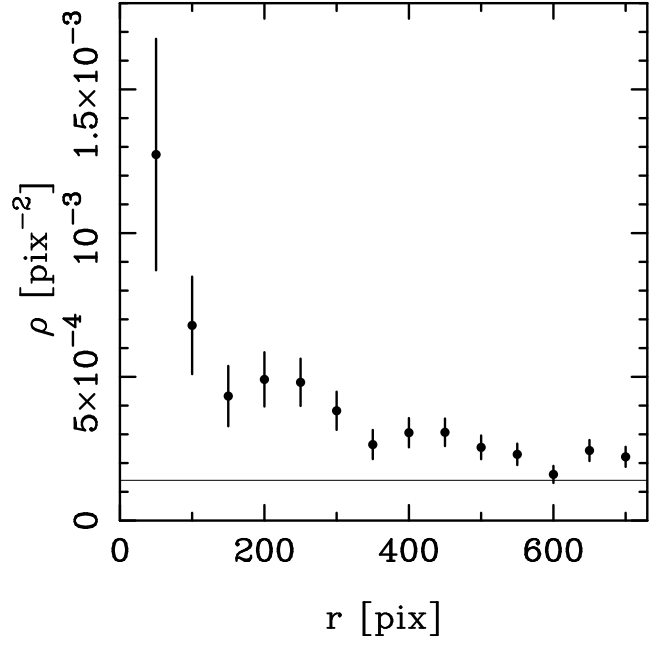}
\caption{Radial variation of the projected stellar density of stars
(filled circles) in the S207 cluster region with center of $\alpha_{\rm
2000} = 04^{\rm h} 19^{\rm m} 42.0^{\rm s}$, $\delta_{\rm 2000} =
+53^\circ 09' 43.9''$.
50\,pixels correspond to $\sim$6$''$.
The error bars represent Poisson errors.  The horizontal solid line
denotes the star density in the sky frame.}  \label{fig:profile_S207}
\end{center}
\end{figure}

In the pseudocolor image of the observed field with MOIRCS
(Fig.~\ref{fig:3col_S207}, bottom), a number of red sources and small 
nebulosities are found along the circular PDR-shell of S207.
Among them, we found a star cluster in the western region of S207 from
the enhancement of stellar density compared to that of the surrounding
area.
Because this enhancement is only seen in the western part of the image,
we used stars only in the western half-frame in the following discussion
of this section (S207 frame).
First, we defined the cluster region. 
We set many circles with 100 pixel ($\sim$12$''$) radius around the
cluster within 600\,pix with 1 pixel step and counted the numbers of
stars included in all circles (3$\pm$2 stars on average).
Among them, we picked up a circle that contains the maximum numbers of
stars (approximately 10) to define the center of the cluster with an
accuracy of $\sim$5$''$:
% center (1528, 1234) ---> (04:19:41.963, +53:09:43.88)
%\textcolor{blue}{($\alpha_{\rm 2000} = 04^{\rm h} 19^{\rm m} 42.0^{\rm
%s}$, $\delta_{\rm 2000} = -53^\circ 09' 43.9''$)}
$\alpha_{\rm 2000} = 04^{\rm h} 19^{\rm m} 42.0^{\rm s}$,
$\delta_{\rm 2000} = +53^\circ 09' 43.9''$. 
Figure~\ref{fig:profile_S207} shows the radial variation of the 
projected stellar density using the stars detected in MOIRCS $K_S$
photometry with $>$5$\sigma$.
The horizontal solid line indicates the density of the entire sky
frame. 
We defined the cluster region with a circle having a radius of 550\,pix
($64''$), where the stellar density is more than that of the entire sky
frame by 3\,$\sigma$. 
In our previous work \citep{Yasui2010}, we defined the cluster region
for S207 using a circle with radius of 200\,pix (23$''$), 
where the lower limit of the stellar density considering three times the
Poisson error is higher than that of the sky frame by 3\,$\sigma$ using
the radial variation with only stars detected in the $K_S$-band
long-exposure image.
%
%from the radial variation with only stars detected in the $K_S$-band
%long-exposure image, considering three times the Poisson errors in each
%bin and those for stellar density of the sky frame.
%
%we estimated the radius of the cluster region at 200\,pix
%from the radial variation of the projected stellar density with only
%stars detected in the $K_S$-band long-exposure image, considering three
%times of the Poisson errors for stellar density in each bin and
%3$\sigma$ errors for stellar density of the sky frame.
%
\color{black}
In this study, we extended the cluster region for better statistics. 

The defined cluster region is shown as the yellow circle in the bottom
figure of Fig.~\ref{fig:3col_S207}. 
The cluster is located near the region where the emission of WISE band 3
(12\,$\mu$m) is very strong (see Fig.~\ref{fig:3col_S207}, top);
%(Fig.~\ref{fig:3col_2MASS_WISE});
%
this combination is often seen in clusters (see \citealt{Koenig2012}).
The cluster radius corresponds to 2.6\,pc and 1.3\,pc with distance of
 $D=8$\,kpc and $D=4$\,kpc, respectively.
 We defined detected point sources in the sky frame as ``field objects''
for a comparison sample of stars in the cluster region.

\subsection{Color--magnitude Diagram} \label{sec:CM}

\begin{figure}[!h]
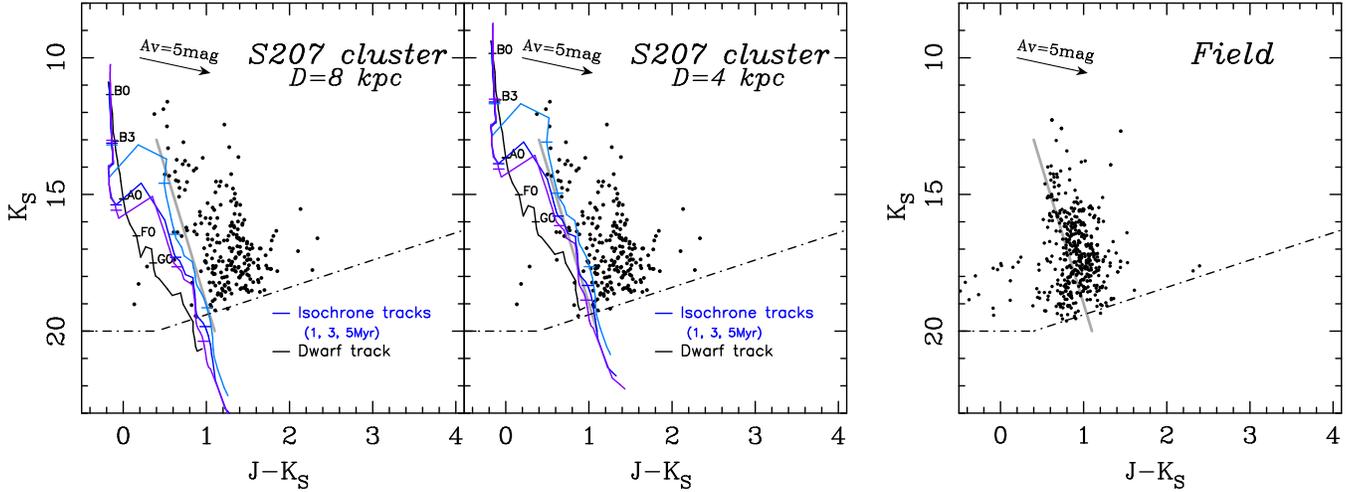

\begin{center}
\includegraphics[height=6.5cm]{fig3a_rev1.eps}
\hspace{1em}
\includegraphics[height=6.5cm]{fig3b_rev1.eps} 
\caption{$(J - K_S)$ vs. $K_S$ color--magnitude diagram {for stars in}
the S207 cluster region ({\it left} and {\it middle}) and the {field
objects} ({\it right}).
Only stars detected with more than 5\,$\sigma$ in both $J$
and $K_S$ bands are plotted.
The arrows show the reddening vectors of $A_V=5$\,mag.
The dot-dashed lines mark the limiting magnitudes (5$\sigma$).
Left and middle: A distance of 8\,kpc is assumed in the left figure,
whereas a distance of 4\,kpc is assumed in the middle figure.
%
%Stars in the cluster region are shown with red filled circles, while all
%the sources outside the cluster region are shown with black dots.
%
The black lines show the dwarf tracks by \citet{Bessell1988} in the
spectral type of O9 to M6 (corresponding mass of
$\sim$0.1--20\,$M_\odot$).
The aqua, blue, and purple lines denote the isochrone models for the age
 of 1, 3, and 5\,Myr old, respectively,
%by \citet{{D'Antona1997},{D'Antona1998}} in the mass range of
% 0.017--3\,$M_\odot$, 
by \citet{{D'Antona1997},{D'Antona1998}} ($0.017 \le M/M_\odot \le 3$),
\citet{Siess2000} ($3 < M/M_\odot \le 7$), and \citet{Lejeune2001} ($7 <
M/M_\odot \le 25$). 
The short horizontal lines are placed on the isochrone models and are
shown with the same colors as the isochrone tracks, which show the
positions of 0.1, 1, 3, and 10\,$M_\odot$.
For convenience, the isochrone models are approximated by the straight
lines, shown as solid gray lines, for estimating the $A_V$ value for
each star.
Right: The solid gray line in the left and middle figures is shown.} 
 \label{fig:colmag_S207}
\end{center}
\end{figure}

We constructed the $J-K_S$ versus $K_S$ color--magnitude diagrams of
detected point sources in the S207 cluster (Fig.~\ref{fig:colmag_S207},
left and middle) and the field objects (Fig.~\ref{fig:colmag_S207},
right).
The dwarf star tracks in the spectral type of O9 to M6 (corresponding
 mass of $\sim$0.1--20\,$M_\odot$) by \citet{Bessell1988} are shown as
 black lines, whereas isochrone models for the ages of 1, 3, and 5\,Myr
 are shown as aqua, blue, and purple lines, respectively.
The isochrone models are by \citet{Lejeune2001} for the mass of $7 <
M/M_\odot \le 25$, by \citet{Siess2000} for the mass of $3 < M/M_\odot 
\le 7$, and by \citet{{D'Antona1997},{D'Antona1998}} for the mass of
$0.017 \le M/M_\odot \le 3$.  
A distance of 8\,kpc is assumed in the left figure of
Fig.~\ref{fig:colmag_S207}, whereas that of 4\,kpc is assumed in the
middle figure of Fig.~\ref{fig:colmag_S207}.
Arrows show the reddening vectors of $A_V = 5$\,mag.

\begin{figure}[!h]
\begin{center}
\includegraphics[width=9cm]{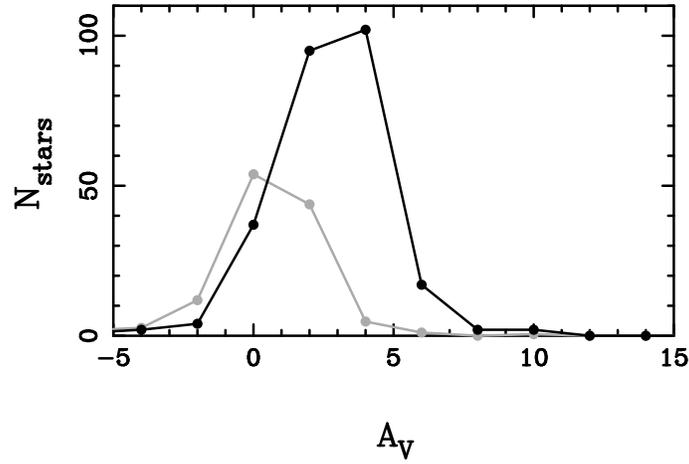} 
\caption{$A_V$ distributions of the stars in the S207 cluster region
({\it black line}) and {the field objects} ({\it gray line}).
 The distribution of field objects is normalized to match 
 with the total area of the cluster region.}
\label{fig:av_S207}
 \end{center}
\end{figure}

In the color--magnitude diagram, the extinction $A_V$ of each star was
estimated from the distance between its location and the isochrone
models along the reddening vector. 
For convenience, the isochrone model is approximated by the straight
line shown as solid gray line.
We then constructed the distributions of the extinction of stars in the
cluster region (black) and {the field objects} (gray) in
Fig.~\ref{fig:av_S207}.
The distribution of field objects is normalized to match with the total
area of the cluster region. 
The resultant distribution of field objects shows a peak at $A_V =
0$\,mag, whereas that for the cluster region shows a peak at the
slightly larger extinction of $A_V \sim 3$\,mag.
Presumably, the difference in the distributions is mainly due to the
cluster members, and the peak $A_V$ value ($\sim$3\,mag) is obtained
from 
%existence of the cluster.  This is probably because there still 
 {existing molecular clouds surrounding the S207 cluster or foreground
 extinction (see detail in Section~\ref{sec:IMF}).}
% (see detail in Section~\ref{sec:SFinS207}).} 
%
The average $A_V$ value of the cluster region is estimated at $A_V =
2.7$\,mag.
Because the differences between the two distributions are not
significant compared to those of the Cloud 2 ($A_V \simeq 3.5$--15\,mag
in \citealt{Yasui2009}), it is difficult to distinguish cluster members
from foreground/background contamination stars in the cluster region
based only on the values of $A_V$.
Therefore, we discuss the properties of this cluster in the following
section by comparing the number count of stars in the cluster region and
{the normalized number counts of field objects}. 
%, which is normalized to match with the total area of the cluster
%regions 
%\color[named]{Gray} (Therefore, we discuss the properties of this
% cluster in the following section by subtracting the number count of
% normalized field objects from that of stars in the cluster region.)
% \color{black}

We placed the short horizontal lines on the isochrone models shown with
the same colors as the isochrone tracks, which show the positions of
0.1, 1, 3, and 10\,$M_\odot$. 
Assuming the average $A_V$ of 3\,mag, the K-band limiting magnitude of
19.0\,mag (10\,$\sigma$) for the age of 1--3\,Myr corresponds to the
mass of $\sim$0.2\,$M_\odot$ and $\lesssim$0.1\,$M_\odot$ in the case of
photometric distance of $D=8$\,kpc and kinematic distance of $D=4$\,kpc,
respectively.
For the age of 5\,Myr, the magnitude corresponds to the mass of
0.5\,$M_\odot$ and 0.09\,$M_\odot$ with distance of 8\,kpc and 4\,kpc, 
respectively.
%in the case of photometric distance and kinematic distance,
%
In any case, the mass detection limit is sufficiently low, down to the
substellar mass, which enables us to estimate the age using KLF
(Section~\ref{sec:KLF}, \ref{sec:Age_Distance}) and derive the disk
fraction with the same criteria as in the solar neighborhood
(Section~\ref{sec:DF}).
Because the most likely age and distance of S207 are estimated at
$\sim$2--3\,Myr and $\sim$4\,kpc, respectively, in
Section~\ref{sec:Age_Distance}, the mass detection limit is then
$\lesssim$0.1\,$M_\odot$.

Note that there are some stars with $J-K_S$ of $\sim$$-$0.5--0.5\,mag in
$K_S$ magnitudes of $\gtrsim$18\,mag, while such stars are not included
in the S207 frame.  Although the scatter in $J-K_S$ colors is often seen
for fainter stars (e.g., Lucas et al. 2008), it appears to be somewhat
large considering the detection limit.
Because the magnitude errors for such stars are found to be relatively
large, $\sim$0.1 mag, there may be some systematic errors 
% due to a small number of frames in J and H bands for the sky frame,
% and 
due to possible unstable sky level at the observation night, when it is
highly humid, despite the the relatively high quality of the photometry
(Section~3.3).
However, because the effects of such stars are very small, the ratio of
only 4\,\% among all field objects, 
%and normalized number of 4.2 (original number of 16).
%However, because the ratio of such stars are negligible, ~***%, 
we use all field objects for a comparison sample of the cluster region
except for estimation of the disk fraction, which is very sensitive to
photometric errors (see Sections~\ref{sec:CC} and \ref{sec:DF}).

\subsection{Color--color Diagram} \label{sec:CC}

In Fig.~\ref{fig:CC_207}, we show the $J-H$ versus $H-K_S$ color--color
diagram for stars in the S207 cluster region (left) and {for the field
stars} (right).
 Only stars with detection of more than 20$\sigma$ for all $JHK_S$ bands
are plotted.
Although the threshold is often set to be 10$\sigma$ or 5$\sigma$
detection, we set the high threshold considering the somewhat high
photometric errors in fainter magnitudes as noted in
Section~\ref{sec:CM}.  
Even with the detection limit, the K-band limiting magnitude of
20$\sigma$ detection ($K_S = 18.4$\,mag) for the age of 1--5\,Myr
corresponds to the mass of $\le$0.8\,$M_\odot$ and $\le$0.3\,$M_\odot$
in the case of photometric distance of $D=8$\,kpc and kinematic distance
of $D=4$\,kpc, respectively.  \color{black}
The dwarf star tracks in the spectral type of late B to M6 in the MKO
system by \citet{Yasui2008} are the solid curves.
The classical T Tauri star (CTTS) loci, originally derived by
\citet{Meyer1997} in the CIT system, are shown as gray lines in the MKO
system \citep{Yasui2008}. 
Arrows show the reddening vectors of $A_V = 5$\,mag.
Stars with circumstellar dust disks are known to show a large $H-K$
color excess (e.g., \citealt{Lada1992}).

The intrinsic $(H-K)$ colors of each star were estimated by dereddening
along the reddening vector to the young star locus in the color--color
diagram (see Fig.~\ref{fig:CC_207}). For convenience, the young star
locus was approximated by the extension of the CTTS locus, and only
stars that are above the CTTS locus were used.
 We constructed intrinsic $H-K$ color distributions for stars in the
cluster region (gray solid line and gray-filled circles) and for field
objects in the sky frame (gray dot-dashed line and gray-filled squares)
in Fig.~\ref{fig:HK0_S207}. 
The distribution of field objects is normalized to match with the total
area of the cluster region.
 The distribution for the S207 cluster, shown by the black thick line
 with black-filled circles, is made by subtracting the counts of
 normalized distribution for field objects from those for the cluster
 region.
Error bars are the uncertainties from Poisson statistics. 
The average $(H-K)_0$ value for stars in the cluster region is estimated
at 0.29\,mag from 99 stars, whereas that for field objects is estimated
at 0.23\,mag from 113 stars (29.38 stars after normalization).  The
average $(H-K)_0$ value for the S207 cluster is estimated at 0.32\,mag
$((0.29 \times 99 - 0.23 \times 29.38) / (99-29.38))$. 
%
%
%
%We also show the difference of two distributions by subtracting the
%distribution for the control field from that for the cluster region,
%shown by black thick line with black-filled circles. 
%
% and cluster members.
%, respectively % and 􏱳0.5 mag 
%typical valueは$\sim$0.2mag, which is consistent
%with average value in the control field (Muench+2002?) 
%
%\color{red} Therefore, there are no significant difference in $(H-K)_0$
%values between the S207 cluster and stars in the control field. 
%
The difference in the average $(H-K)_0$ between {the stars in the S207
cluster and the field objects in the sky frame (0.09\,mag)} can be
attributed to thermal emissions from the circumstellar disks of the
cluster members.
Assuming that disk emissions appear in the $K$ but not in the $H$ band,
the disk color excess of the S207 cluster members in the K band, $\Delta
K_{\rm disk}$, is equal to 0.09\,mag.

\subsection{K-band Luminosity Function (KLF)} \label{sec:KLF}

We constructed the K-band luminosity function (KLF) for the S207
cluster.  Because it is difficult to identify cluster members from $A_V$
values (cf. \citealt{Yasui2009}), as discussed in Section~\ref{sec:CM},
we construct KLF by subtracting the normalized star counts of field
objects from those for the stars in the cluster region.
In Fig.~\ref{fig:KLF_obs}, we show three KLFs: (1) KLF for stars in the
cluster region shown by gray solid lines and gray-filled circles
(cluster region KLF);
(2) KLF for {field objects with normalized star counts to match with the
 total area of the cluster regions, shown by} gray dot-dashed lines and
 gray filled-squares (field KLF);
and (3) KLF for background-subtracted star counts (S207 KLF) {shown by} 
black thick lines and black-filled circles.
The number counts of KLFs generally increase in the fainter magnitude
bins. 
However, the number counts of the cluster region KLF decrease in the
faintest bin, and the S207 KLF has peaks at $K=18$\,mag bin, which
generally corresponds to the peak of IMF.

It should be noted that the detection completeness of stars with
$<$10\,$\sigma$ detection is less than one, whereas that of the brighter
stars is almost one (see \citealt{Yasui2008} and \citealt{Minowa2005}).
Considering the 10\,$\sigma$ detection magnitude of $K=19.0$\,mag 
for both the S207 frame and the sky frame, completeness should be
$\sim$1 in the magnitude bins of $K=13$--18\,mag.
However, in the $K=19$\,mag bin, which includes stars with
$K=18.5$--19.5\,mag, not all stars may be perfectly detected owing to
the small detection completeness of stars with $K=19.0$--19.5\,mag
corresponding to the 5\,$\sigma$--10\,$\sigma$ detection. {The number
counts in the bin of the cluster region KLF and the field KLF are 59 and
41.7, respectively.}
The number of stars with $K=19$--19.5\,mag in the cluster region KLF is
counted as 21{, while that in the field KLF is 20.0}.
Because the completeness of stars with 5\,$\sigma$ detection is
estimated at $\sim$0.7 using \citet{Yasui2008} (also see
\citealt{Minowa2005}), the actual numbers of stars with
$K=19$--19.5\,mag are 30 $(21/0.7)$ {and 28.6 $(20.0/0.7)$} at most {for 
the cluster region KLF and the field KLF, respectively}.
By combining this with the number of stars with $K=18.5$--19\,mag (38
{for the cluster regiond KLF and 21.7 for the field KLF}), the upper
limit of the number count in $K=19$\,mag bin for the cluster region KLF
becomes 68 $(=30+ 38)$, shown as gray open circles and gray dashed
lines{, that for the field KLF becomes 50.3 $(=28.6+21.7)$, shown as
gray open squares and gray dashed lines}.
Therefore, the resultant upper and lower limits of the star count in the
S207 KLF are estimated at 26.3 $(= 68 - 41.7)$, shown as the black open
circles and dashed lines, and 8.7 $(= 59 - 50.3)$, shown as dotted
lines, respectively. 
As a result, the S207 KLF would have a peak at $K=18$\,mag bin.

%%%%%%%%%%%%%%%%%%%%%%%%%%%%%%%%%%%%%%%%%%%%%%%%%%%%%%%%%%%%%%%%%%%%%%%%%%%%%%%
%%%%%%%%%%%%%%%%%%%%%%%%%%%%%%%%%%%%%%%%%%%%%%%%%%%%%%%%%%%%%%%%%%%%%%%%%%%%%%%

\section{Discussion}

\subsection{Age and distance of S207} \label{sec:Age_Distance}

% label{sec:Age}

{KLFs of different ages are known to have different peak magnitudes and
slopes, fainter peak magnitudes and less steeper slopes} with increasing
age \citep{Muench2000}. 
By comparing observed and model KLFs, the age of the young
clusters can be roughly estimated with an uncertainty of $\pm$1\,Myr
(\citealt{{Yasui2006},{Yasui2008}}). 
We constructed model KLFs in the same way as our previous work (see
Section~4 in \citealt{Yasui2006}) with the assumed distance of the
cluster that underlies typical IMFs and mass--luminosity relations.
We used the Trapezium IMF \citep{Muench2002}, which is considered the
most reliable IMF for young clusters (e.g., \citealt{LadaLada2003}), as
discussed in \citet{Yasui2008}.
Because the maximum mass of the stars in S207 seems much larger compared
to that of the clusters in our previous work ($\sim$3\,$M_\odot$;
\citealt{{Yasui2006},{Yasui2008}}),
we used additional isochrone models by \citet{Lejeune2001} for the mass
of $7 < M/M_\odot \le 25$ and by \citet{Siess2000} for the mass of $3 <
M/M_\odot \le 7$, in addition to the model for the lower mass ($0.017
\le M/M_\odot \le 3$) by \citet{{D'Antona1997},{D'Antona1998}}, which is
used in \citet{{Yasui2006},{Yasui2008}}.
We constructed model KLFs with ages 0.5, 1, 2, 3, and 5\,Myr assuming
kinematic distance of $D=8$\,kpc (Fig.~\ref{fig:KLF_fit}, {\it left})
and photometric distance of $D=4$\,kpc (Fig.~\ref{fig:KLF_fit}, {\it
right}).
We also considered the $A_V$ and $\Delta K_{\rm excess}$ estimated in
Sections~\ref{sec:CM} and \ref{sec:CC}.

We performed a chi-squared test between the S207 KLF (black line in
Fig.~\ref{fig:KLF_fit}) and model KLFs for each age and distance with
magnitudes of $K=12.5$--18.5\,mag.
The results of the best fit for fixed ages are shown as colored lines in
Fig.~\ref{fig:KLF_fit}.
The obtained chi-square values are not very important because the
constructed model KLFs are extremely simplified (e.g., age spread is not
considered). However, the values can be used to check whether the
assumed ages are likely or not.
In the case of $D=8$\,kpc (Fig.~\ref{fig:KLF_fit}, {\it left}), the most
likely age is approximately 0.5\,Myr or 1\,Myr with chi-square values of
$\sim$30 (26.9 for the age of 0.5\,Myr and 31.6 for the age of 1\,Myr;
35.8, 33.4, and 37.3 for 2\,Myr, 3\,Myr, and 5\,Myr, respectively). 
However, the ages are implausible based on the very high chi-square
values, which are much less than the 5\,\% {and 1\,\%} confidence level
in the degree of 5 (11.07 {and 15.09, respectively}). 
In the case of $D=4$\,kpc (Fig.~\ref{fig:KLF_fit}, {\it right}), the
ages of 2\,Myr, 3\,Myr, or 5\,Myr seem most reliable based on the
chi-square value of 
$\sim$13 (13.2, 13.1, and 13.2 for the age of 2\,Myr, 3\,Myr, and
5\,Myr, respectively; 51.1 and 51.7 for 0.5\,Myr and 1\,Myr,
respectively), which is between 5\,\% and 1\,\% confidence limits.
However, in the case of 5\,Myr, the model KLF does not match the
observed KLF in the faintest $K=19$\,mag bin at all\footnote{In case 
that the faintest $K=19$\,mag bin is also included in the KLF fitting
for the distance of 4\,kpc, obtained chi-square values for the age of
2\,Myr and 3\,Myr (16.4 and 13.2, respectively) are close to 5\,\% and
1\,\% confidence limit in the degree of 6 (14.44 and 16.81,
respectively), while that for the age of 5\,Myr (34.9) is much more than
the limits.}.
Therefore, the age of 2--3\,Myr in the case of distance $D=4$\,kpc seems 
most likely for the S207 cluster.

The KLF fitting results suggest that the distance of the S207 cluster
is consistent with the kinematic distance and that S207 is located in
the outer arm, as described in Section~\ref{sec:properties}.
Because circular Galactic rotation is suggested for the outer arm
\citep[e.g.,][]{Hachisuka2015}, the kinematic distance is likely to be
correct,
whereas that in the Perseus arm is incorrect because of the noncircular
motion \citep{Xu2006}.
The photometric distance may have large uncertainty because it is
estimated from an O-type star, for which the luminosity class is
difficult to determine.
In addition, the estimate is only from one star.
Actually, the photometric distance from an A-type star is much different
($D \simeq 2$\,kpc), as discussed in Section~\ref{sec:properties}.
For star-forming \ion{H}{2} regions, the re-examined distances are often
found to differ from previously estimated distances by more than a
factor of 2 \citep[e.g.,][]{Russeil2007}.

\subsection{Implication for IMF} \label{sec:IMF}

In Section~\ref{sec:CC}, the extinction of stars in the S207 cluster is
estimated as small as $A_V \sim 3$\,mag, which should include foreground
extinction.
Because $A_V \sim 3$\,mag is expected in the direction of S207 from the
three-dimensional extinction map of the Galactic plane \citep{Sale2014},
the extinction from the intracluster should be very small.
Moreover, the non-detection of CO by \citet{Blitz1982} with sensitivity
limit of 0.5\,K
suggests that the 3\,$\sigma$ upper limit of the H$_2$ column
density\footnote{The telescopes at Bell Telescope Laboratories (BTL) and
Millimeter Wave Observatory (MWO) were used for the S207 observation in
\citet{Blitz1982}.
The column density is estimated considering a main beam efficiency of
89\,\% for BTL \citep{Bally1987} and $\sim$80\,\% for MWO
\citep{Magnani1985},
and the velocity resolution of 0.65\,km\,s$^{-1}$ in \citet{Blitz1982}.}
is $\sim$3--4$\times$10$^{20}$\,cm$^{-2}$, which corresponds to {$A_V < 
0.4$\,mag}.
The estimated extinction values suggest that the molecular cloud around
the S207 cluster almost dissipated.
Considering that star-forming clusters experience embedded 
phase for only 2--3\,Myr \citep{LadaLada2003}, the small extinction of
the S207 cluster suggests that the age of the cluster is $\sim$2--3\,Myr,
which is consistent with the age estimated from the KLF fitting.

In the KLF fitting (Section~\ref{sec:Age_Distance}), typical IMF is
assumed for estimating the age of the S207 cluster.
In contrast, if we take the above rough age estimate as an independent
information, the fitting results with model KLFs suggest that the IMF of
the S207 cluster, which is located in very low-metallicity environments,
can be approximated by the typical IMFs of the solar neighborhood
($\sim$0\,dex) for mass range $>$0.1\,$M_\odot$.
Because the KLF shape is very sensitive to that of IMF
\citep{{Yasui2006},{Yasui2008ASPC}}, 
the reasonably good fit of the observed KLF down to the peak at
$m_K = 18$\,mag ensures the universality of IMF.

\subsection{Disk fraction} \label{sec:DF}

\begin{figure}[h]
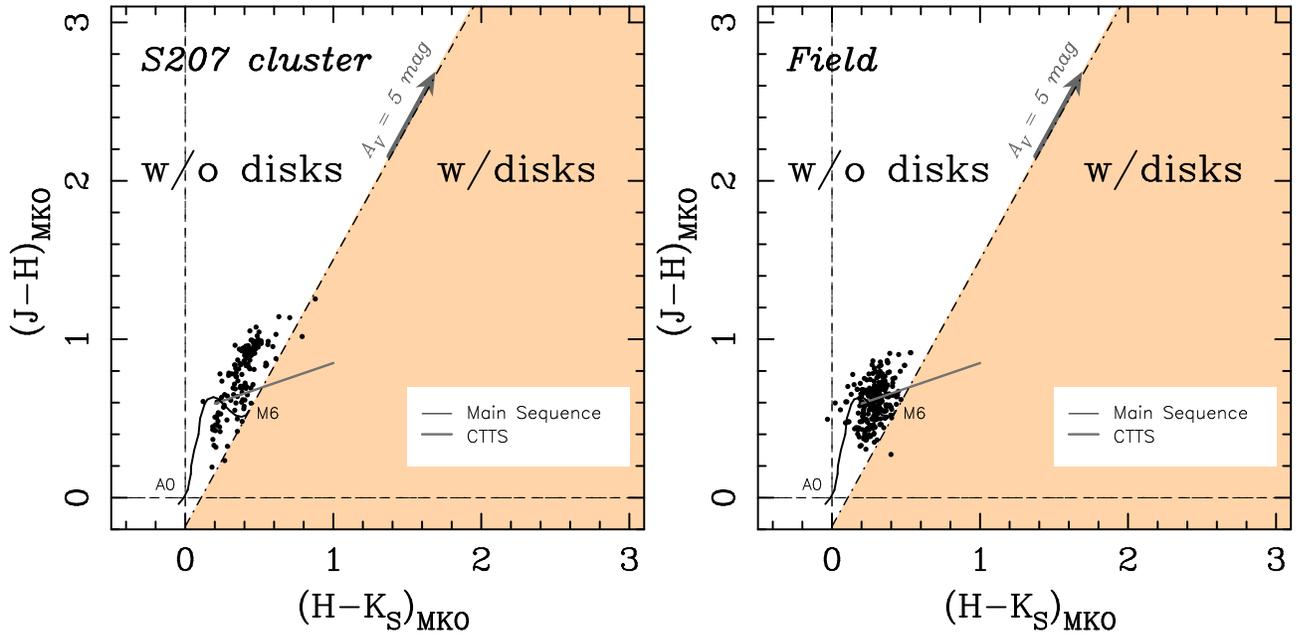

\begin{center}
\hspace{2em}
%
% cd /Users/chikako/WORK/MOIRCS2006/S207/plfile2014/rev2015
% ./CC_S207SL_ch2_rev2015.pl 1
% \includegraphics[width=8.5cm]{/Users/chikako/WORK/MOIRCS2006/S207/plfile2014/rev2015/CC_S207SL_ch2_rev2015_20sig.eps}
\includegraphics[width=8.5cm]{fig5a_rev1.eps}
%
% CC_S207SL_ch2_rev2015_skyALL.pl
%\includegraphics[width=8.5cm]{/Users/chikako/WORK/MOIRCS2006/S207/plfile2014/rev2015/CC_S207SL_ch2_rev2015_skyALL_20sig.eps}
\includegraphics[width=8.5cm]{fig5b_rev1.eps}
\caption{$(H-K_S)$ vs. $(J - H)$ color--color diagrams of the stars in
the S207 cluster region (left) and the {field objects} (right).
 Only stars detected with more than 20\,$\sigma$ in all
$JHK_S$ bands are plotted. \color{black}
The solid curves in the lower left portion of the diagram are the loci
of points corresponding to the unreddened main-sequence stars.  The
dot-dashed lines that intersect the main-sequence curves at maximum
$H-K_S$ values (M6 point on the curve) and are parallel to the reddening
vector are the borders between stars with and without circumstellar
disks. The classical T Tauri star (CTTS) loci are shown by the gray
lines (see details in \citealt{Yasui2008}).}
 \label{fig:CC_207}
\end{center}
\end{figure}

The ratio of stars with protoplanetary disks in young clusters, the disk
fraction, is one of the most fundamental parameters in planet formation
\citep{{Haisch2001ApJL},{LadaLada2003}}.
On the JHK color--color diagram, stars without circumstellar disks are
seen as main-sequence stars reddened with extinction, whereas stars with
circumstellar disks are seen in ``the disk-excess region,'' which is the
orange highlighted region to the right of the dot-dashed line in
Fig.~\ref{fig:CC_207} because of thermal emissions from the hot dust
disk with temperature of $\sim$1500\,K.
The dot-dashed line intersecting the dwarf star curve at maximum $H-K_S$
values (M6 point on the curve) and is parallel to the reddening vector
is the border between stars with and without circumstellar disks (see
details in \citealt{Yasui2009}).
The disk fraction for stars with more than {20$\sigma$} detection for
all $JHK_S$ bands in the cluster region is estimated at {2\,\% $(3 /
138)$} from Fig.~\ref{fig:CC_207} (left). 
However, because some stars in the cluster region are indeed background
or foreground stars, we have to subtract such contaminations (e.g.,
\citealt{Haisch2000AJ}).
The disk fraction for {field objects in the sky frame} is estimated at
{0.4\,\% $(1/247)$} from Fig.~\ref{fig:CC_207} (right). 
As a result, the final disk fraction of the S207 cluster is estimated at
4$\pm$2\,\% $((9-0.26) / (138 - 64.22))$ with the number of field
objects normalized to match with the total area of the cluster regions. 
In our previous work \citep{Yasui2010}, we estimated the disk fraction
for the cluster at 5$\pm$5\,\%: 5$\pm$4\,\% for stars in the cluster
region, which is a small region with radius $r=200$\,pix, and 5\,\% for
stars in the control field, which is the area located by more than
500\,pix from the center of the cluster region.
In this study, we defined the larger area ($r=550$\,pix; see
Section~\ref{sec:S207cluster}) as the cluster region {to estimate disk
fraction with higher S/N.}
In addition, to avoid possible cluster members, we defined the control
field in the sky frame, which is well apart from the nebulosity of S207.
In any case, the discrepancy between the two results was within the
uncertainties range.

\begin{figure}[!h]
\begin{center}
\includegraphics[width=8.5cm]{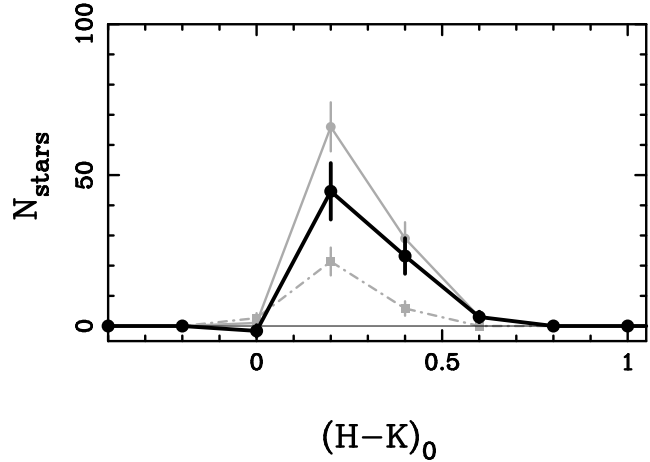}
\caption{$(H-K)_0$ distributions for stars in the S207 cluster region
{(gray solid line with gray-filled circles)} and that for field objects
(gray dot-dashed line with gray-filled squares). 
 The distribution of the field objects is normalized to
match with the total area of the cluster region. 
 The distribution of the S207 cluster, shown by the black thick line
 with black-filled circles, is made by subtracting the distribution for
 the control field region from that for the cluster region.} 
\label{fig:HK0_S207}
 \end{center}
\end{figure}

\begin{figure}[!h]
\begin{center}
\includegraphics[width=8.5cm]{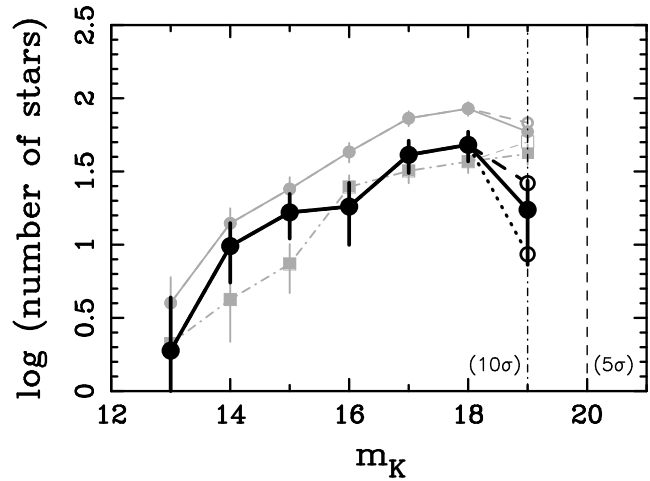}
\caption {The raw KLF for stars in the cluster region (cluster region
KLF) and {for field objects} (field KLF) are shown by the gray solid
line with gray-filled circles and gray dot-dashed line with gray-filled
squares, respectively.
 The star counts of field objects are normalized to match
 with the total area of the cluster region. 
The KLF for the S207 cluster (S207 KLFs), shown by the black thick line
with black-filled circles, is made by subtracting the {normalized} star
counts in the field KLF from {star counts} in the cluster region KLF.
Error bars are the uncertainties from Poisson statistics.
The upper limit of the counts in $K=19$\,mag bin for the
 cluster region KLF is shown by the gray dashed line with the gray open
 circle considering the detection completeness, while that for the
 controld field KLF is shown by the gray dashed lines with the gray open
 square.
The resultant upper and lower limits of the counts in $K=19$\,mag bin
 for the S207 KLF are shown with the black dashed line with the black
 open circle and black dotted line with the black circle, respectively.
The vertical dot-dashed line and vertical dashed line show the limiting
 magnitudes of the 10\,$\sigma$ detection (19.0\,mag) and 5\,$\sigma$
 detection (20.0\,mag), respectively.}
\label{fig:KLF_obs}
\end{center}
\end{figure}

\begin{figure}[!h]
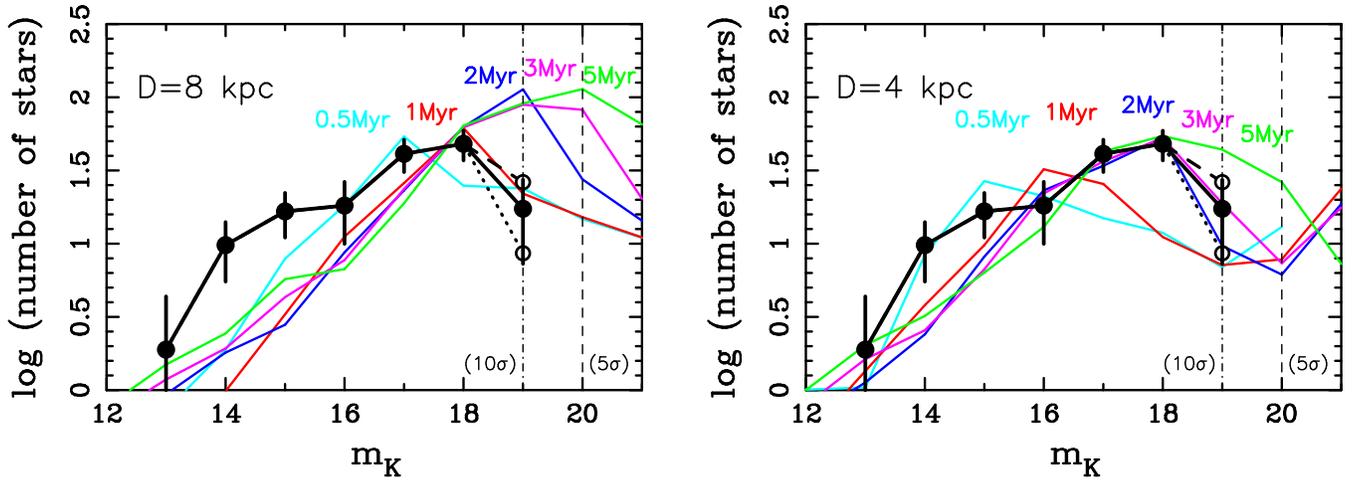

\begin{center}
\includegraphics[width=8.5cm]{fig8a_rev1.eps}
\hspace{2em}
\includegraphics[width=8.5cm]{fig8b_rev1.eps}
\caption{Comparison of the S207 KLFs (black lines from Fig.8) with model 
KLFs of various ages (colored lines).
The dashed lines with the open circles {and the dotted lines with the
 open circles} represent the upper {and lower} limits of the counts for
 $K = 19$\,mag bin for the S207 cluster considering the detection
 completeness. 
Two cases for the distance are assumed: photometric distance of
$D=8$\,kpc (left) and kinematic distance of $D=4$\,kpc (right).
The aqua, red, blue, magenta, and green lines represent model KLFs of
0.5, 1, 2, 3, and 5 Myr, respectively.}
\label{fig:KLF_fit}
\end{center}
\end{figure}

\begin{figure}[!h]
\begin{center}
\includegraphics[width=8.5cm]{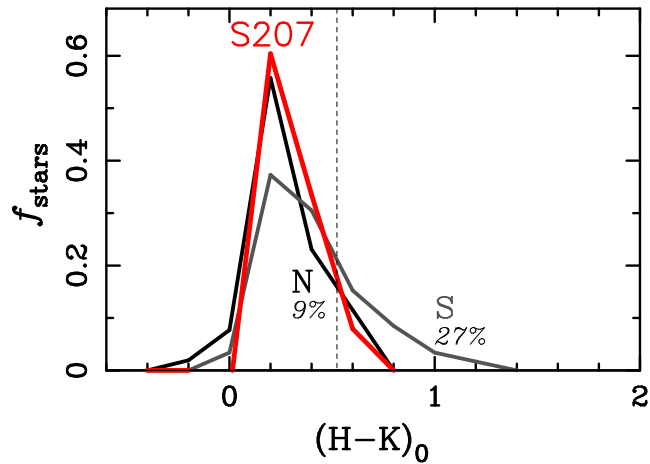}
\caption {Comparison of intrinsic $H-K$ color
distributions. The fractions of stars ($f_{\rm stars}$) per each
intrinsic color bin ($(H-K)_0$) 
%is plotted.
%
%Top: $(H-K)_0$ distributions 
for clusters in low-metallicity environments, S207, Cloud 2-N (labeled
 with ``N''), and Cloud 2-S (labeled with ``S'') are plotted. 
%
%Disk fractions of the Cloud 2-N and Cloud 2-S clusters are 
%9\,\% and 27\,\%, respectively.
%
%
%
 The dashed line shows the borderline for estimating the disk fraction
 in the MKO system.} 
%. Note that the border of the $(H-K)_0$ value for estimating
% the disk fraction (dashed line) differs slightly for different filter
% system, MKO and CIT tystem for the top and bottom figure,
% respectively.}
%
\label{fig:HK0_disk}
\end{center}
\end{figure}

In Figure~\ref{fig:HK0_disk}, we show the fraction of stars ($f_{\rm
stars}$) per each intrinsic $(H-K)$ color bin ($(H-K)_0$) for the S207
cluster (red), which is from black line in Fig.~\ref{fig:HK0_S207}), and
those for other young clusters in the low-metallicity environments,
Cloud 2-N (black) and Cloud 2-S (gray) with estimated disk fraction of
9\,\% and 27\,\%, respectively \citep{Yasui2009}.
The dashed line shows the borderline for estimating the disk fraction in
the MKO system\footnote{Although $(H-K)_0 = 0.43$\,mag was shown for the
borderline in the MKO system in \citet{Yasui2009}, we found that the
correct borderline is $(H-K)_0 = 0.52$\,mag.}.
The distribution is known to be bluer and sharper with lower disk
fraction for nearby young clusters (see the bottom panel of Fig.~7 in
\citealt{Yasui2009}), which is also the case for the clusters in the
low-metallicity environments \citep{Yasui2009}.
%The distribution is known to be bluer and sharper with lower disk
%fractions (see the bottom panel of Fig.~7 in \citealt{Yasui2009}).
%
The distribution of the S207 cluster appears to be sharp with dispersion
of $\Delta (H-K)_0 \sim 0.4$\,mag, which resembles the distribution of 
the Cloud 2-N cluster.
This is consistent with that disk fraction of the S207 is low,
$<$10\,\%.

NIR disk fractions are known to have high values ($\sim$60\,\%) for very
young clusters but decrease with increasing age.
In the timescale of $\sim$10\,Myr, the fraction reaches $\sim$5--10\,\%
 (\citealt{Lada1999}; \citealt{Hillenbrand2005}; \citealt{Yasui2010};
 see the red line in the left figure of Fig~5 in \citealt{Yasui2014}).
Although NIR disk fractions are generally slightly lower than MIR disk
 fractions, which are based on ground $L$-band observations and space
 MIR observations, the characteristics are totally identical.
As suggested in \citet{Yasui2010}, the derived disk fraction for the
 S207 cluster (4$\pm$2\,\%) is lower than that for clusters in the solar
 neighborhood with identical age to the S207 cluster ($\sim$30\,\% for
 $\sim$2--3\,Myr; \citealt{Yasui2010}).
The lower disk fraction in low-metallicity environments suggests that
the disk lifetime in low-metallicity environments is quite short, as
discussed in \citet{Yasui2009}.

%%%%%%%%%%%%%%%%%%%%%%%%%%%%%%%%%%%%%%%%%%%%%%%%%%%%%%%%%%%%%%%%%%%%%%%%%%%%%%%
\acknowledgments

%We thank the anonymous referee for careful reading and thoughtful
%suggestions that improved this paper. 

This work was supported by JSPS KAKENHI Grant Number 26800094. 
%KAKENHI. \\
%
We thank the Subaru support staff, in particular, the MOIRCS support
astronomer Ichi Tanaka.
We also thank Chihiro Tokoku for helpful discussions on the
observation. 

%Kiso observatory staff. 

%%%%%%%%%%%%%%%%%%%%%%%%%%%%%%%%%%%%%%%%%%%%%%%%%%%%%%%%%%%%%%%%%%%%%%%%%%%%%%%

%%% Tab1 %%%
\begin{table*}[h]
\caption{Properties of S207.}\label{tab:targets}
\begin{center}
\begin{tabular}{llcccccccc}
\hline
\hline
Name & Sh 2-207 \\
Galactic longitude (deg) &   151.1905 (1) \\
Galactic latitude (deg) &  $+$2.1256 (1) \\
R.A. (J2000.0) & 04 19 49.6  (1) \\
Dec. (J2000.0) & +53 09 29  (1) \\
Photometric heliocentric distance$^{\rm a}$ (kpc) & 7.6 (2), 8.6 (3) \\
Adopted photometric heliocentric distance (kpc) & 8 \\
%($D_{\rm phot}$) &\\
Photometric Galactocentric distance$^{\rm b}$ (kpc) & $\simeq$15.5\\
Kinematic heliocentric distance$^{\rm c}$ (kpc) &  4.3 (4), 3.4 (5) \\ 
Adopted kinematic heliocentric distance (kpc) &  4\\
%($D_{\rm kin}$) & \\
Kinematic Galactocentric distance$^{\rm d}$ (kpc) &  $\simeq$12\\ 
Oxygen abundance $12 + \log {\rm (O/H)}$ 
& 7.96 {(6, 7)} \\
Metallicity [O/H] (dex)$^{\rm e}$ & $-$0.8 \\
% $7.96^{+0.17}_{-0.27}$ \\ 
%
\hline
\end{tabular}
\end{center}
%
%{{\small N{\scriptsize OTES.---}\\
{{\small {\bf Notes.} \\
%{N{\scriptsize OTES.---}
%{N{\scriptsize OTES.---}
%
%
$^{\rm a}$References are shown in the parenthesis. \\
$^{\rm b}$Assumed {the solar Galactocentric distance} $R_\odot =
 8.0$\,kpc. References are shown in the parenthesis. \\
%  the Galactocentric distance of the Sun
$^{\rm c}$References are shown in the parenthesis. \\
$^{\rm d}$Assumed the solar Galactocentric distance $R_\odot =
  8.0$\,kpc. References are shown in the parenthesis. \\
$^{\rm e}$Assumed the solar abundance of $12+ \log {\rm (O/H)} = 8.73$
 \citep{Asplund2009}. \\ 
%
% estimated with $\log {\rm [O/H]} = 8.7$ \citep{Asplund2005} [dex]. \\
%
%
%
%$^{\rm a}$ Galactic radius. 
%$^{\rm b}$  References for the ages are shown in the parenthesis.\\
%$^{\rm b}$ Mass detection limit of the data.\\
%
{\small \bf References. }
%{\scriptsize EFERENCES. ---} %(a) \citet{Smartt1996}, (b)
%R{\scriptsize EFERENCES. ---} %(a) \citet{Smartt1996}, (b)
% \citep{Digel1994}, (c)  \citet{Stil2001}, (d) \citet{Lubowich2004}, 
%
(1) SIMBAD \citep{Wenger2000}, {(2)} \citet{Moffat1979}, {(3)}
\citet{Chini1984}, {(4)} \citet{Fich1990}, {(5)} \citet{Pismis1991},
{(6)} \citet{Caplan2000}, and {(7) } \citet{Rudolph2006}.}}
%(j) \citet{Lahulla1985}.}}
%(k) \citet{Vilchez1996}.
%
\end{table*}
%\end{landscape}

%%%%%%%%%%%%%%%%%%%%%%%%%%%%%%%%%%%%%%%%%%%%%%%%%%%%%%%%%%%%%%%%%%%%%%%%%%%%%%
%%%% Tab 2 %%%
%\begin{table*}[!h]
\begin{table*}
\caption{Summary of MOIRCS observations.} \label{tab:LOG}
\begin{center}
\begin{tabular}{lccccccc}
\hline
\hline
Modes & Band & $t_{\rm total}$ & $t$ & Coadd & $N_{\rm total}$
 & Seeing \\
%NDR &
& & (1) & (2) & (3) & (4) \\
%\hline
%Cloud 2 & Sep 3, 2006 & $J$ & 1200 & 120 & 1 & 10 & $0.''.32$ & P\\
%Cloud 2 & Sep 3, 2006 & $H$ & 1200 & 20 & 6 & 10  & $0.''.37$ & P\\
%Cloud 2 & Sep 3, 2006 & $K$ & 1200 & 40 & 3 & 10  & $0.''.35$ & P\\
\hline  %& $1''.2$
$J$-long &  $J$ & 360 (360) & 120 & 1 & 3 (3) & $1''.1$ \\ 
$H$-long &  $H$ & 480 (360) & 20 & 6 & 4 (3) & $0''.9$ \\
$K_S$-long &  $K_S$ & 960 (720) & 30 & 4  & 8 (6) & $0''.9$ \\  
$J$-short &  $J$ & 52 (39) & 13 & 1 & 4 (3) & $1''.2$ \\
$H$-short &  $H$ & 52 (39) & 13 & 1 & 4 (3) & $1''.2$\\
$K_S$-short &  $K_S$ & 52 (39) & 13 & 1 & 4 (3) & $1''.2$\\
%date	積分時間 ( sec x coadd x dither)	seeing	comments
%long	2006.11.07 (2006B)	J: 120x1x4, H: 20x6x4, Ks: 30x4x4 (2セット）	~0.9" (7.5 pix)	◎
%short	2006.11.07 (2006B)	J: 13x1x4, H: 13x1x4, Ks: 13x1x4	~ 2" (10 pix)	◎
\hline 
\end{tabular}
\end{center}
{{\small {\bf Notes.} \\
%{N{\scriptsize OTES.---}
Col. (1): Total exposure time [s]. The values for the sky frames are
 shown in parentheses. 
% The time for the sky frames are shown in parentheses.
%
Col. (2): Single-exposure time [s].
Col. (3): Number of coadd. 
%The number for the sky frame is shown in parentheses. 
%
Col. (4): Total number of frames. The values for the sky frames are
 shown in parentheses.
%Col.(4): Number of non-destructive readouts.
%
%
%Col.(5): P: photometric, H: high humidity, PS: photometric (terminated
% with snow), C: cirrus.
}}
\end{table*}

%%%%%%%%%%%%%%%%%%%%%%%%%%%%%%%%%%%%%%%%%%%%%%%%%%%%%%%%%%%%%%%%%%%%%%%%%%%%%%%
%%% Tab 3 %%%
\begin{table*}[!h]
%
%\captionsetup{labelfont=bf,font={sf,small}}
\caption{Limiting magnitudes of long-exposure images for MOIRCS
observations.} \label{tab:limit}
\begin{center}
\begin{tabular}{lccc} 
\hline
\hline
Frame & $J$ (10$\sigma$/5$\sigma$) & $H$  (10$\sigma$/5$\sigma$) 
& $K_S$ (10$\sigma$/5$\sigma$) \\ %& Cluster\\
%Target & $J$ & $H$ & $K_S$ & Cluster\\
%\hline
%Cloud 2 (ch1) & 22.2/- & 21.3/- & 21.0/- & Cloud 2-S\\
%Cloud 2 (ch2) & 22.3/- & 21.7/- & 21.2/- & Cloud 2-N\\
\hline
%Ks limit (10/5σ): 18.5/19.3 mag	H limit(10/5σ): 19.2/20.1 mag	J limit(10/5σ): 19.8/20.5 mag
Cluster & 19.6/20.4 & 19.1/20.0 & 19.0/20.0 \\
Sky & 19.6/20.4 & 19.0/19.8 & 19.0/20.0\\ 
%\hline
%Sh 2-208 & 20.1/21.0 & 19.0/19.9 & 18.5/19.8 \\
%J(10/5 sigma) = 20.1, 21.0 mag
%H(10/5 sigma) = 19.0, 19.9 mag
%Ks(10/5 sigma) = 18.5, 19.8 mag
%\hline
% ch1 
%Sh 2-209 (ch1) & 21.9/22.8 & 20.9/21.7 & 20.5/21.6 & sub-cluster \\
% ch2
%Sh 2-209 (ch2) & 22.0/22.9 & 21.0/21.9 & 20.7/21.2 & main cluster \\
\hline
\end{tabular}
\end{center}
\end{table*}

%%% fig2 %%%
% xpa frame
% xpa file S207_Ks_chip2_long.fits
%%% xpa file dS207obj_Ks_SkyS_ch12_2015Feb.fits
%%% cat S207_IRAS_B9.reg | xpaset ds9 regions -format ds9 -system wcs
% 
% 全て log-98%？
% xpa file dS207obj_Ks_SkyS_ch12_WCS_2015Feb.fits
% xpa file dS207obj_H_SkyS_ch12_WCS_2015Feb.fits
% xpa file dS207obj_J_SkyS_ch12_WCS_2015Feb.fits
% 
% xpaset -p ds9 regions delete all 
% cat S207_IRAS_B9_0305.reg | xpaset ds9 regions -format ds9 -system wcs
% cat S207_IRAS_B9_rot270_0305.reg | xpaset ds9 regions -format ds9 -system wcs
%
% /Users/chikako/WORK/MOIRCS2006/S207/txtfile2014/APphotS207sky_2015_ch2.txt
% dS207obj_Ks_SkyS_ch12_2015Feb.fits,dS207obj_H_SkyS_ch12_3col_2015Feb.fits,dS207obj_J_SkyS_ch12_3col_2015Feb.fits 
% IRAS: (04 19 49.6, +53 09 29), B9星は、星の上にプロットする

%%% fig2 %%%

%%% fig3 %%%   color-magnitude diagram

%%% fig4 %%%

%%% fig5 %%% color-color

%%% fig6 %%%  (H-K)0

%%%%%%%%%%%%%%%%%%%%%%%%%%%%%%%%%%%%%%%%%%%%

%%% fig7 %%%  

%%% fig8 %%% KLF fitting

%%% fig9 %%%  

\end{document}